\documentclass[twocolumn,amssymb,nobibnotes,aps,prl, reprint,letterpaper,preprintnumbers,amsmath,superscriptaddress,showpacs,floatfix]{revtex4-1}
\usepackage{graphicx,xcolor}% Include figure files [dvips]
\usepackage{dcolumn}% Align table columns on decimal point
\usepackage{bm}% bold math
\usepackage{graphics}
\usepackage{float}
\usepackage{epsfig}
\usepackage{multirow}
\usepackage{amsmath}
\usepackage{tabularx}
\usepackage{mathrsfs}
\usepackage[pass,paperwidth=8.5in,paperheight=11in]{geometry}
\usepackage{lineno}
\usepackage[linktocpage=true]{hyperref}
\usepackage{hyperref}
\usepackage{comment}
\usepackage{soul}
\usepackage{notes2bib}
\hypersetup{
  pdfnewwindow=true, colorlinks=true,
  linkcolor=blue, anchorcolor=blue,
  citecolor=blue, filecolor=blue,
  menucolor=blue, urlcolor=blue}

\begin{document}
\title{Capping effects on spin and charge excitations in parent and superconducting Nd$_{1-x}$Sr$_{x}$NiO$_2$}

\author{S. Fan}
\email{sfan1@bnl.gov}
\affiliation{National Synchrotron Light Source II, Brookhaven National Laboratory, Upton, NY 11973, USA.}

\author{H. LaBollita}
\affiliation{Department of Physics, Arizona State University, Tempe, AZ, USA.}

\author{Q. Gao}
\affiliation{Institute of Physics and Beijing National Laboratory for Condensed Matter Physics, Chinese Academy of Sciences, Beijing 100190, China}

\author{N. Khan}
\affiliation{Department of Physics and Astronomy, Rutgers University, Piscataway, NJ 08854, USA.}

\author{Y. Gu} 
\affiliation{National Synchrotron Light Source II, Brookhaven National Laboratory, Upton, NY 11973, USA.}

\author{T. Kim} 
\affiliation{National Synchrotron Light Source II, Brookhaven National Laboratory, Upton, NY 11973, USA.}

\author{J. Li} 
\affiliation{National Synchrotron Light Source II, Brookhaven National Laboratory, Upton, NY 11973, USA.}

\author{V. Bhartiya} 
\affiliation{National Synchrotron Light Source II, Brookhaven National Laboratory, Upton, NY 11973, USA.}

\author{Y. Li}
\affiliation{National Laboratory of Solid State Microstructures, Jiangsu Key Laboratory of Artificial Functional Materials, College of Engineering and Applied Sciences, Nanjing University, Nanjing, China 210046}

\author{W. Sun}
\affiliation{National Laboratory of Solid State Microstructures, Jiangsu Key Laboratory of Artificial Functional Materials, College of Engineering and Applied Sciences, Nanjing University, Nanjing, China 210046}

\author{J. Yang}
\affiliation{National Laboratory of Solid State Microstructures, Jiangsu Key Laboratory of Artificial Functional Materials, College of Engineering and Applied Sciences, Nanjing University, Nanjing, China 210046}

\author{S. Yan}
\affiliation{National Laboratory of Solid State Microstructures, Jiangsu Key Laboratory of Artificial Functional Materials, College of Engineering and Applied Sciences, Nanjing University, Nanjing, China 210046}

\author{A. Barbour}
\affiliation{National Synchrotron Light Source II, Brookhaven National Laboratory, Upton, NY 11973, USA.}

\author{X. Zhou}
\affiliation{Institute of Physics and Beijing National Laboratory for Condensed Matter Physics, Chinese Academy of Sciences, Beijing 100190, China}
\affiliation{Songshan Lake Materials Laboratory, Dongguan 523808, China}

\author{A. Cano}
\affiliation{CNRS, Université Grenoble Alpes, Institut Néel, 38042 Grenoble, France}

\author{F. Bernardini}
\affiliation{Dipartimento di Fisica, Università di Cagliari, IT-09042 Monserrato, Italy}

\author{Y. Nie}
\affiliation{National Laboratory of Solid State Microstructures, Jiangsu Key Laboratory of Artificial Functional Materials, College of Engineering and Applied Sciences, Nanjing University, Nanjing, China 210046}

\author{Z. Zhu}
\affiliation{Institute of Physics and Beijing National Laboratory for Condensed Matter Physics, Chinese Academy of Sciences, Beijing 100190, China}
\affiliation{Songshan Lake Materials Laboratory, Dongguan 523808, China}

\author{V. Bisogni}
\affiliation{National Synchrotron Light Source II, Brookhaven National Laboratory, Upton, NY 11973, USA.}

\author{C. Mazzoli}
\affiliation{National Synchrotron Light Source II, Brookhaven National Laboratory, Upton, NY 11973, USA.}

\author{A. S. Botana}
\affiliation{Department of Physics, Arizona State University, Tempe, AZ, USA.}

\author{J. Pelliciari}
\email{pelliciari@bnl.gov}
\affiliation{National Synchrotron Light Source II, Brookhaven National Laboratory, Upton, NY 11973, USA.}

\begin{abstract}
%The discovery of superconductivity in infinite layer nickelates (Nd$_{1-x}$Sr$_x$NiO$_2$) has attracted a lot of attention. So far, 
Superconductivity in infinite layer nickelates Nd$_{1-x}$Sr$_x$NiO$_2$ has so far been achieved only in thin films raising questions on the role of substrates and interfaces. Given the challenges associated with their synthesis it is imperative to identify their intrinsic properties. %On this end, capping with SrTiO$_3$ affects the interface structure possibly playing an active role on the electronic and magnetic properties of nickelates. 
%However, only few spectroscopic studies specifically addressed this topic. 
We use Resonant Inelastic X-ray Scattering (RIXS) to investigate the influence of the SrTiO$_3$ capping layer on the excitations of Nd$_{1-x}$Sr$_x$NiO$_2$ ($x$ = 0 and 0.2). Spin excitations are observed in parent and 20\% doped Nd$_{1-x}$Sr$_x$NiO$_2$ regardless of capping, proving that magnetism is intrinsic to infinite-layer nickelates and appears in a significant fraction of their phase diagram. In parent and superconducting Nd$_{1-x}$Sr$_x$NiO$_2$, the spin excitations are slightly hardened in capped samples compared to the non-capped ones. Additionally, a weaker Ni - Nd charge transfer peak at $\sim$ 0.6 eV suggests that the hybridization between Ni 3$d$ and Nd 5$d$ orbitals is reduced in capped samples. From our data, capping induces only minimal differences in Nd$_{1-x}$Sr$_x$NiO$_2$ and we phenomenologically discuss these differences based on the reconstruction of the SrTiO$_3$ - NdNiO$_2$ interface and other mechanisms such as crystalline disorder.

\end{abstract}

\maketitle
\begin{figure}[t] % use p to make a pure float page
\centering
\includegraphics[width=3.43in]{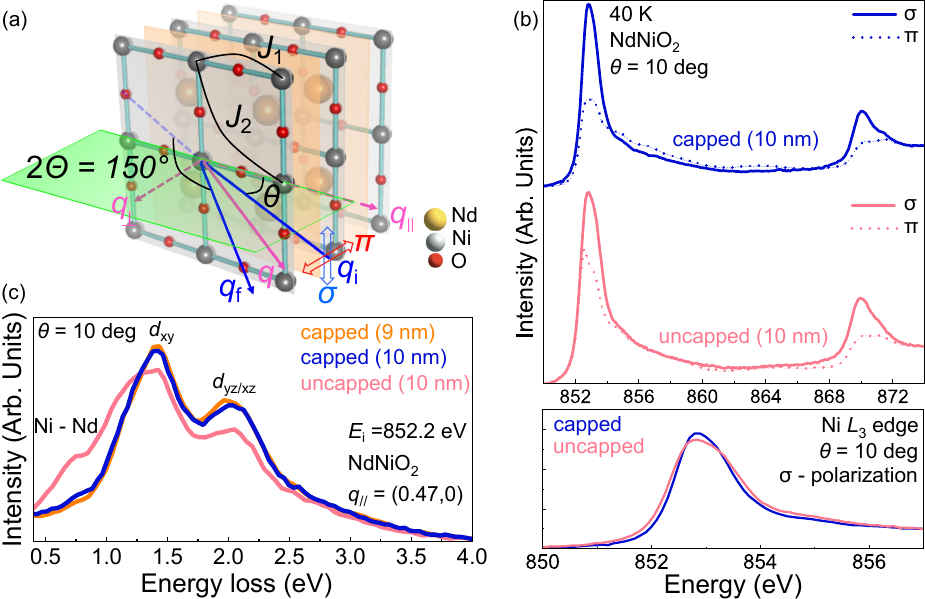}
 \caption{(a) Geometry of XAS and RIXS. $\vec{q_i}$ and $\vec{q_f}$ represent the wave-vectors of incident and scattered X-rays. The grazing angle $\theta$ is defined as the angle between the incident x-ray and the sample surface. The 2$\Theta$ scattering angle (angle between $\vec{q_i}$ and $\vec{q_f}$) is fixed at 150 degrees to maximize the momentum transfer.  $\vec{q_{\parallel}}$ and $\vec{q_{\perp}}$ refer to that the momentum transfer direction is parallel or perpendicular to the Ni-O bonds, respectively. %$J_1$ and $J_2$ refer to the first and second nearest neighbour exchange constant, respectively. 
 (b) XAS linear dichroism of the NdNiO$_2$ - cap and NdNiO$_2$ - uncap (both 10 nm) at Ni $L_3$/$L_2$ edges. The grazing angle $\theta$ is 10 deg. %Data points filled with or without color correspond to the $\sigma$- or $\pi$- polarization directions, respectively.
 Bottom: comparison of XAS of the NdNiO$_2$ - cap/uncap at the Ni $L_3$ edge with $\sigma$-polarization. (c) RIXS spectra of the NdNiO$_2$ - cap/uncap with $\theta$ = 10 deg, $\sigma$-polarization, and $E_i$ = 852.2 eV.} %Error bars of the raw data are smaller than the marker's size.}
\label{RIXS geometry}
\end{figure}

The search for cuprate analog materials has led to the discovery of superconductivity in infinite layer (RNiO$_2$), quintuple-layer (Nd$_6$Ni$_5$O$_{12}$), and bilayer (La$_3$Ni$_2$O$_7$) nickelates \cite{Li2019,Pan2021,Sun2023}. Infinite layer nickelates RNiO$_2$ (R=Nd, Pr, and La) are weak insulators composed of NiO$_2$ planes lacking apical oxygens separated by rare-earth (R) layers [Fig.~\ref{RIXS geometry}(a)] \cite{Li_absence_2020,lee_aspects_2020,wang_synthesis_2020,puphal_topotactic_nodate,Li2019}. In undoped RNiO$_2$, nickel has a nominal Ni$^{1+}$ oxidation state, a $d^9$ electronic configuration, and $d_{x^2-y^2}$ orbital polarization similar to parent cuprates, and once hole-doped, RNiO$_2$ displays superconductivity \cite{Li2019}. Resonant Inelastic X-ray Scattering (RIXS) detected dispersive spin excitations with $J$ $\approx$ 64 meV \cite{Lu2021} in capped NdNiO$_2$ significantly lower in energy than in cuprates \cite{Braicovich2009,Tacon2011,Peng2018}, whereas, in uncapped NdNiO$_2$, a peak at $\approx$ (1/3, 0, $L$) r.~l.~u. concomitant with the absence of spin excitations was reported and a charge ordered ground state was inferred \cite{Lu2021,Rossi2022,Krieger2022,Pelliciari2023,Parzyck2023,Ren2023,Tam2022}. From a synthesis perspective, superconductivity of RNiO$_2$ has so far been detected only in thin films, a key distinction from cuprates \cite{Li_absence_2020,lee_aspects_2020,wang_synthesis_2020,puphal_topotactic_nodate,Bernardini2022} which raises questions on the role of substrate, capping, and interface. 

In the past, multiple phenomena have been observed at the interface of oxide thin films as due to the combination of polar/non-polar surfaces, charge transfer, phonon coupling, confinement and strain, among others \cite{Huang2017,Wang2012,Lee2018b,hwang_emergent_2012,soumyanarayanan_emergent_2016,pelliciari_evolution_2021, pelliciari_tuning_2021,Fan2020}.~These interface modifications can affect a large portion of the film due to either the limited thickness or long-range interactions \cite{Fan2020,catalan_progress_2008,Mundy2014,Ohtomo2004}.~Electron-energy loss spectroscopy (EELS) and scanning transmission electron microscopy (STEM)
%\textcolor{blue}{For infinite layer nickelates, prior studies show the superconductivity is observed for different thickness (from 5.1 to 17 nm) samples, and thicker films tend to show higher T$_C$. \cite{Nomura2021,Zeng2021} This brings the question of understanding the role of capping layer since the reconstruction of electronic structures are naturally expected at the interface. \cite{Nomura2021,He2020,Geisler2020,Zhang2020} However, the relation between those interface effects and superconductivity is not trivial and still under debate. \cite{Ji2021,He2020,Botana2021}} 
revealed a thickness-dependent atomic and electronic reconstruction at the NdNiO$_2$ - SrTiO$_3$ interface, the effects of which can propagate to the inner NiO$_2$ layers \cite{Yang2023,Goodge2023}. %~The intermediate Nd(Ti,Ni)O$_3$ layer \textcolor{blue}{is formed at the SrTiO$_3$ - NdNiO$_2$ interface} which quenches the formation of a two-dimensional electron gas at the interface, modifying the electronic structure of the whole nickelate film \cite{Goodge2023}. %This evidence emphasizes the importance of interfacial effects and the SrTiO$_3$ capping layer. 
Relevant to the presence or absence of capping the following evidence can be extracted:~(i) capping modifies the distribution of apical oxygen during the chemical reduction \cite{Parzyck2023,Yang2023,Raji2023}; (ii) a $q$ = (1/3, 0, $L$) peak (which was assigned as the charge density wave) is detected in uncapped nickelates with weak or absent spin excitations \cite{Tam2022,Rossi2022,Krieger2022,Ren2023,Raji2023}; and (iii) superconductivity is observed in both capped and uncapped cases with some sample-specific differences in T$_C$ \cite{Krieger2022,Geisler2020,He2020,Zhang2020,Bernardini2020,Rossi2021,Nomura2021,Zeng2021,Botana2021,Ji2021}.
However, experimental information on the capping layer and its repercussion on the electronic and magnetic excitations have still to be explored using high-end spectroscopies.

Here we combine X-ray Absorption Spectroscopy (XAS) and high-resolution RIXS to investigate the electronic excitations of Nd$_{1-x}$Sr$_{x}$NiO$_2$ ($x$ = 0, 0.2) as a function of Sr doping and in the presence/absence of the SrTiO$_3$ capping layer. High quality Nd$_{1-x}$Sr$_{x}$NiO$_2$ display strong orbital polarization and spin excitations are detected regardless of capping. Strontium doping softens the spin excitations due to magnetic dilution achieved through hole doping on the nickel site \cite{Lu2021,Rossi2021}.~Furthermore, spin excitation energies of the capped parent and superconducting compounds are slightly increased compared to the uncapped counterparts.~A suppressed charge transfer peak at $\sim$ 0.6 eV in capped samples indicates that capping reduces the hybridization between Ni 3$d$ and Nd 5$d$ orbitals. This evidence proves that magnetism is an intrinsic property of infinite layer nickelates and that capping has only minor effects on the spin and charge excitations in infinite-layer nickelate films. %and these modifications are mostly at the interface \cite{Goodge2023,Yang2023}.}

%can be rationalized with the atomic and electronic reconstructions due to the polar discontinuity at the SrTiO$_3$-NdNiO$_2$ interface as indicated by our Density Functional Theory (DFT) calculations and RIXS data on NdNiO$_2$ of different thicknesses, ultimately confirming the active role of capping.

We studied:~(i) NdNiO$_2$ with and without capping (NdNiO$_2$ - cap/uncap, 10 nm thick) and (ii) Nd$_{0.8}$Sr$_{0.2}$NiO$_2$ with and without capping (Nd$_{0.8}$Sr$_{0.2}$NiO$_2$ - cap/uncap). The thickness of the SrTiO$_3$ capping layer is 3 nm for both parent and doped films. All films are grown on SrTiO$_3$ \footnote{See Supplemental Information, which includes Refs \citenum{Gao2021prepare,Li2021prepare,Fumagalli2019b,Lin2020,KresseVasp1,KresseVasp2,BlahaWien2k,Perdew1996generalized}, for additional information about the experimental methods, $dd$ excitation assignments, fitting details, and details of the Density functional theory calculations}. A 9 nm NdNiO$_2$ - cap is also measured for comparison. The high sample quality is confirmed by resonant x-ray diffraction (See End Matter), electric transport, and X-ray absorption measurements at the Ni-$L$ and O-$K$ edges.~The T$_C$ of the doped samples is 11 and 15 K for  Nd$_{0.8}$Sr$_{0.2}$NiO$_2$ - uncap and  Nd$_{0.8}$Sr$_{0.2}$NiO$_2$ - cap, respectively (Supplementary Fig. 2). We perform XAS at the Ni $L_3$/$L_2$ edges in total electron yield (TEY) and RIXS at the Ni $L_3$ edge with $\sigma$- and $\pi$-polarization using the experimental geometry sketched in Fig.~\ref{RIXS geometry}(a). At $\theta$ = 10 degree, the $\sigma$- or $\pi$- x-ray polarization refers to the electric field vector of the light being along the in-plane Ni-O bond direction or almost along the out-of-plane direction, respectively.  

Figure \ref{RIXS geometry}(b) displays the XAS linear dichroism of NdNiO$_2$ - cap and NdNiO$_2$ - uncap. The XAS corroborates the high quality of our samples as they display a single peak in $\sigma$-polarization (bottom panel) and strong linear dichroism ($\Delta I$) regardless of capping \cite{Zeng2024}. This implies an in-plane orientation of
the unoccupied Ni 3$d_{x^2 - y^2}$ orbitals \cite{Rossi2021,Goodge2021}. The XAS dichroism is sensitive to the presence of spurious phases \cite{Parzyck2023,Raji2023} because the excess apical oxygen in the partially reduced films deplete the anisotropic polarization of the in- and out-of-plane Ni 3$d$ orbitals. This indicates that our samples have minimal spurious signals from secondary phases \cite{Parzyck2023}.
The peak in $\sigma$-polarization ($\approx$852.3 eV) corresponds to the Ni 2$p^63d^9$ $\rightarrow$ 2$p^53d^{10}$ transition related to Ni$^{1+}$ valence state. In $\pi$-polarization, a weak shoulder emerges around 853 eV for uncapped NdNiO$_2$ and is ascribed to the Ni$^{2+}$ state originating from self-doping induced by Ni-Nd orbital hybridization \cite{Lu2021,Hepting2020}. Since the Ni-Nd hybridization is mainly along the out-of-plane direction, the Ni$^{2+}$ is more prominent in the $\pi$-polarized XAS. This is confirmed by the RIXS spectra displayed in Fig.~\ref{RIXS geometry}(c) where the Nd 5$d$ - Ni 3$d$ charge transfer peak at $\approx$ 0.6 eV  is moderately suppressed in the NdNiO$_2$ - cap. Similar capping effects are also consistently observed in the XAS and RIXS spectra of the Nd$_{0.8}$Sr$_{0.2}$NiO$_2$ (Supplementary Fig.~3).

\begin{figure}[t!]
\centering
\includegraphics[width=3.5in]{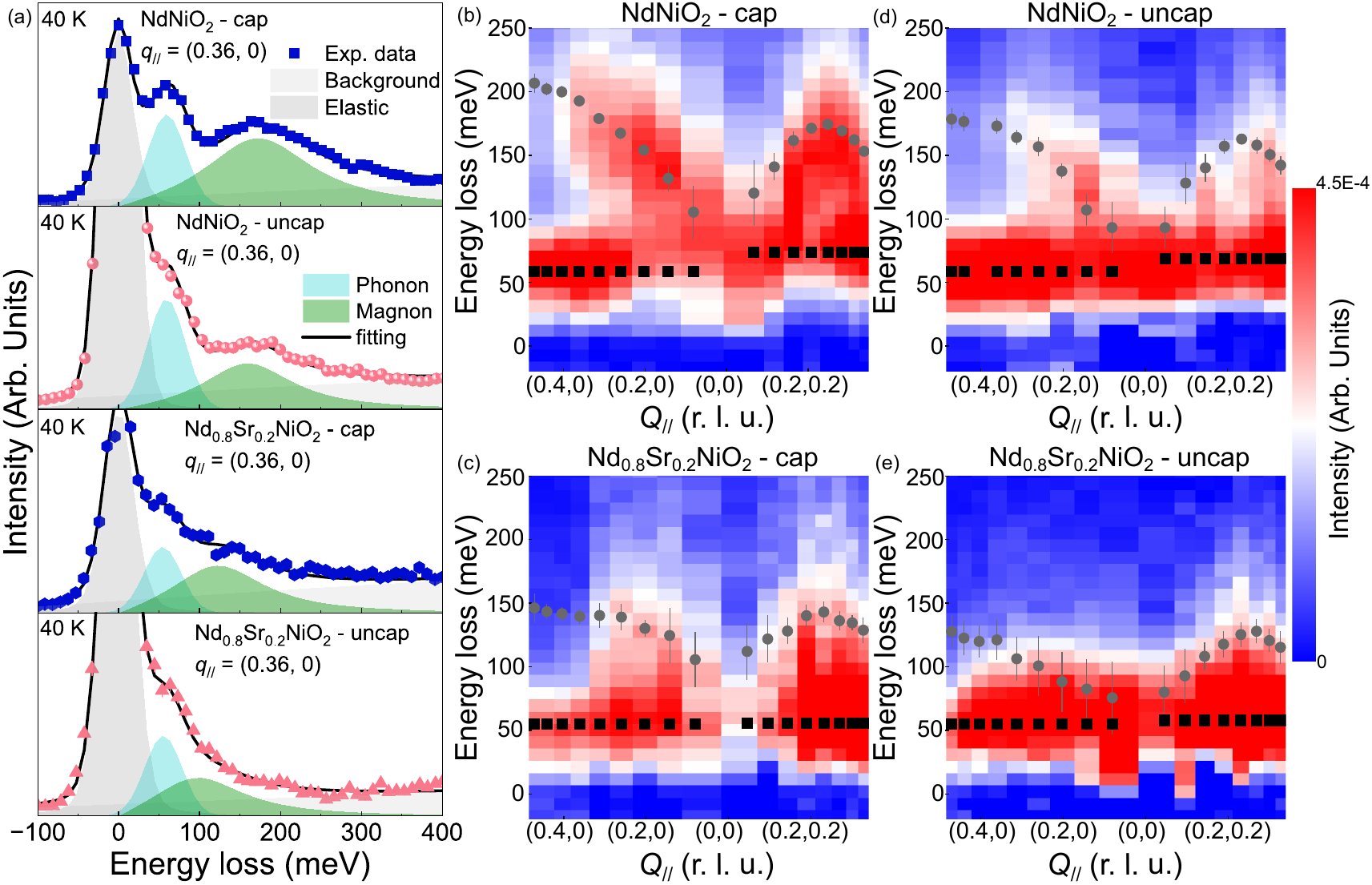}
\caption{(a) RIXS spectra for NdNiO$_2$ - cap, NdNiO$_2$ - uncap, Nd$_{0.8}$Sr$_{0.2}$NiO$_2$ - cap and Nd$_{0.8}$Sr$_{0.2}$NiO$_2$ - uncap at 40 K and $q_{\parallel}$ = (0.36, 0) r.~l.~u.~in $\sigma$-polarization. The spectra are fitted using three peaks.~Light grey represents the background. (b,d) RIXS intensity map vs.~energy loss and in-plane momentum transfer along ($h$, 0) and ($h$, $h$) of the NdNiO$_2$ - cap/uncap, respectively. (c,e) Similar RIXS intensity map as (b,d) but for the Nd$_{0.8}$Sr$_{0.2}$NiO$_2$ - cap/uncap, respectively. Grey and black data points correspond to the phonon and spin excitation dispersions extracted from the fitting.}
\label{LowE excitations}
\end{figure}

Figure \ref{LowE excitations}(a) shows RIXS spectra at low-energy loss ($-100 \rightarrow 400$ meV) in $\sigma$-polarization for all the samples at $q_{\parallel}$ = (0.36, 0) r.~l.~u.. Besides the elastic line, a phonon at $\approx$ 60 meV and a dispersive spin excitation at 100 - 200 meV are clearly identified \cite{Lu2021}. We fit the data with a Gaussian peak for the elastic and phonon~while the spin excitations are fitted via a damped harmonic oscillator \cite{Lu2021,Peng2018} (see Supplementary Information section IV for details). The background is fitted by a broad antisymmetric Lorentzian function capturing the decay from Ni $dd$ excitations. After the removal of the elastic peak, the RIXS data are summarized in Fig. \ref{LowE excitations}(b-e) for ($h$, 0) and ($h$, $h$). The dispersive color bands above 100 meV in NdNiO$_2$ - cap/uncap provide robust evidence on the presence of spin excitations in NdNiO$_2$ irrespective of capping. This observation is at odds with Ref. \citenum{Krieger2022}, where spin excitations are absent for NdNiO$_2$ - uncap. The inconsistency is likely due to distinct sample quality as the XAS look significantly different and we could not observe any resonant peak at $q$ = (1/3, 0, $L$) (see `End Matter') \cite{Parzyck2023,Raji2023,Krieger2022}.
The spin excitations persist in hole-doped samples and their bandwidth softens and intensities are suppressed due to the reduction of exchange bonds induced by hole doping on the nickel sites [Fig. \ref{LowE excitations} (c,e)]. This doping trend seems to be intrinsic of the samples proving the magnetic dilution effect due to hole doping regardless of capping.

\begin{figure}[t!]
\centering
\includegraphics[width=3.5in]{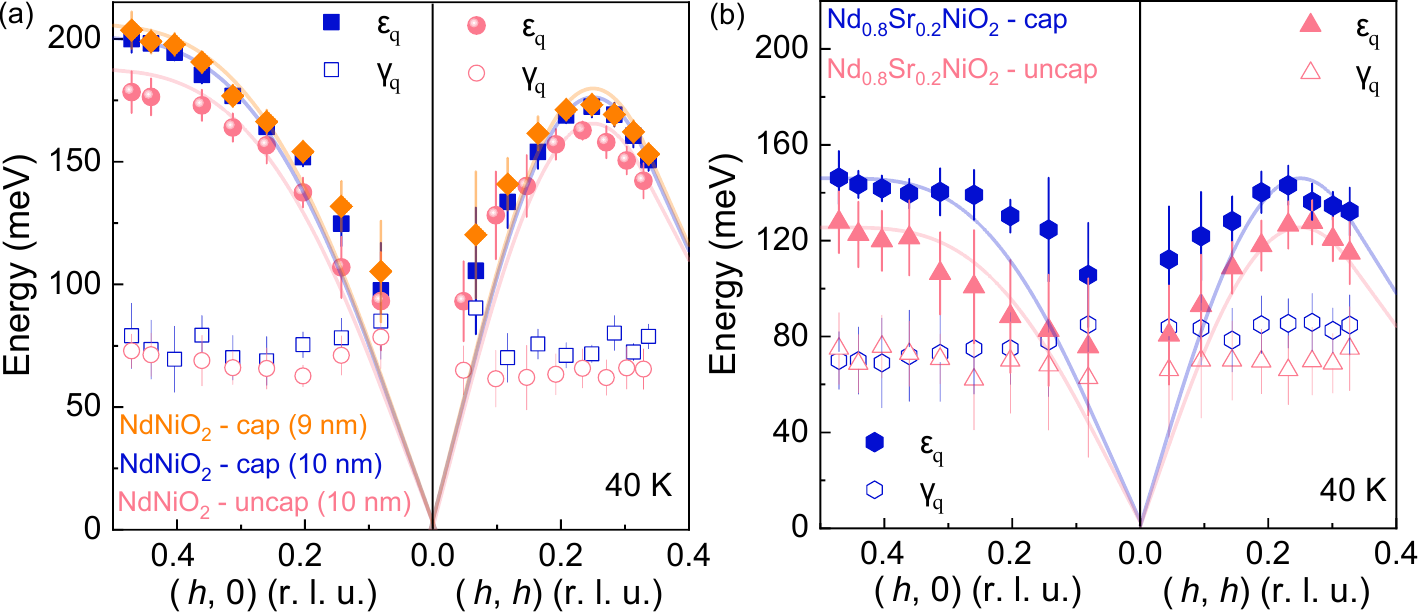}
\caption{ (a,b) Fitted un-damped spin excitation energy $\epsilon_q$ and damping factor $\gamma_q$ as a function of in-plane momentum transfer along ($h$, 0) and ($h$, $h$) of NdNiO$_2$ - cap/uncap (a) and superconducting Nd$_{0.8}$Sr$_{0.2}$NiO$_2$ - cap/uncap (b). The curves in light colors are the spin excitation dispersion from linear spin wave theory. The data of the 9 nm NdNiO$_2$ - cap is also added for comparison.}
%(c) Magnon spectral weight vs.~momentum transfer along ($h$, 0) and ($h$, $h$) for capped and uncapped Nd$_{1-x}$Sr$_x$NiO$_2$ ($x$ = 0 and 0.2). The data are obtained by integrating the fitted spin excitation peak along the energy-loss axis. (d) Total spectral weight of the spin excitation for NdNiO$_2$ - cap (9 and 10 nm) (samples 1,2), NdNiO$_2$ - uncap (10 nm) (sample 3), Nd$_{0.8}$Sr$_{0.2}$NiO$_2$ - cap (sample 4), and Nd$_{0.8}$Sr$_{0.2}$NiO$_2$ - uncap (sample 5), renormalized to the NdNiO$_2$ - cap (10 nm). Error bars are estimated by considering the phonon intensity variation and multiple fittings (see appendix).} 
\label{spin dispersions}
\end{figure}

Figure \ref{spin dispersions}(a) compares the spin excitations' energy of NdNiO$_2$ - cap and NdNiO$_2$ - uncap. 
To quantify the spin dispersion we use linear spin wave theory for the spin 1/2 square-lattice Heisenberg antiferromagnet \cite{Lu2021,Coldea2001}:

\begin{equation}
    H = J_1\sum_{i,j}S_i\cdot S_j + J_2\sum_{i,i'}S_i\cdot S_{i'}
\end{equation}
where $S_i$, $S_j$, and $S_i'$ are spins at site $i$, nearest-neighbor sites $j$, and the next nearest-neighbor sites $i'$, respectively. $J_1$ and $J_2$ are the nearest and second-nearest exchange constants [Fig.~\ref{RIXS geometry}(a)] \cite{Lu2021}. %The exchange coupling constants extracted from the fitting are summarized in Table \ref{Summary of nickelates}.
We obtained $J_1$ $\approx$ 72.8 $\pm$ 4.3 meV and $J_2$ $\approx$ -5.8 $\pm$ 2.0 meV in the NdNiO$_2$ - cap (10 nm), consistent with previous studies \cite{Lu2021} and for the NdNiO$_2$ - uncap, $J_1$ $\approx$ 67.0 $\pm$ 4.0 and $J_2$ $\approx$ -5.2 $\pm$ 2.0 meV. The spin excitation energy of NdNiO$_2$ - cap is systematically higher than that of the uncapped case at all the significant momentum points. We believe that this hardening is intrinsic given that the spin excitation energies are nearly identical for 9 and 10 nm thick NdNiO$_2$ - cap (which we used to identify the sample-to-sample variations). Besides the value of $Js$, the capping-induced hardening of the spin excitations can be more directly identified in the raw RIXS spectra at high momentum (see End Matter).~A similar behavior of the spin excitations hardening in presence of capping is also detected in Nd$_{0.8}$Sr$_{0.2}$NiO$_2$ [Fig.~\ref{spin dispersions}(b)]. In doped films, $J_2$ is set to zero to avoid overfitting due to the reduced bandwith of the spin excitations.

%Figure \ref{spin dispersions}(c) displays the spectral weight of NdNiO$_2$ - cap, NdNiO$_2$ - uncap, Nd$_{0.8}$Sr$_{0.2}$NiO$_2$ - cap, and Nd$_{0.8}$Sr$_{0.2}$NiO$_2$ - uncap as function of momentum. %The data points are obtained by integrating the area of the spin excitations along the energy loss axis. 

%Capping renormalizes the intensity over the momentum space with a notable enhancement of spectral weight at 0.2 $\le$ $h$ $\le$ 0.4 r.~l.~u.. Figure \ref{spin dispersions}(d) displays the sum at all momenta of the integrated intensity of each sample. This quantity can be associated with the fluctuating magnetic moment (upon kinematical limits). 
%\textcolor{blue}{The fluctuating magnetic moment in infinite-layer nickelates is increased by approximately $\approx$ 20\%  NdNiO$_2$ -cap, comparable to the hole doping effects.} The spectral weight of the capped sample gains extra $\sim$ 15\% when the thickness of the NdNiO$_2$ layer reduces to 9 nm \textcolor{blue}{indicative of an increase of the relative weight of the interfaces over the bulk of the film}.

\begin{figure*}[t!]
\centering
\includegraphics[width= 5.0in]{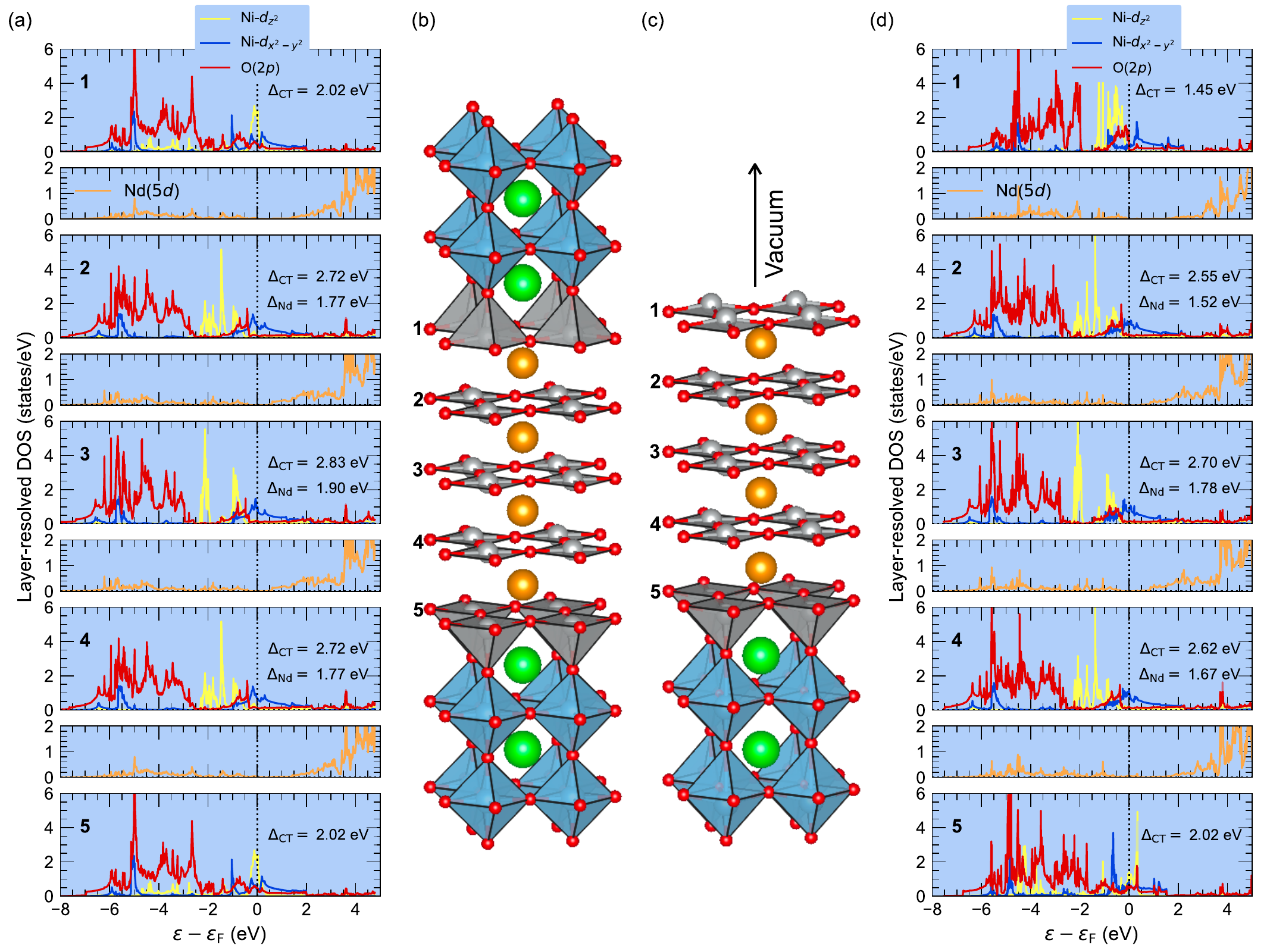}
\caption{Summary of the ``capping" vs.~``uncapping'' effects on the electronic structure of SrO interface for NdNiO$_2$. (a) Layer- and atom-resolved density of states (DOS) within the NdNiO$_2$ block in the ``capped" NdNiO$_2$ with focus on the Nd($5d$), O($2p$), and Ni-$e_{g}$ states. (b) Crystal structure of NdNiO$_2$/SrTiO$_3$ heterostructure with SrO interface which refers to a ``capped'' NdNiO$_2$. (c) Same as (b) for the ``uncapped'' case. (d) Same as (a) for the ``uncapped'' case.}
\label{DFT calculation}
\end{figure*}

Our data highlights that magnetism is intrinsic to Nd$_{1-x}$Sr$_x$NiO$_2$ and that the SrTiO$_3$ capping layer has overall minor effects on the spin and charge excitations. %and it is confined to the interface. 
Importantly, the strong XAS dichroism, absence of any pre-edge peak at the O-$K$ edge (see End Matter), lack of the resonant $q$ = (1/3, 0, $L$) peak (see End Matter), and presence of similar spin excitations in our capped and uncapped films demonstrate that the quality of our capped and uncapped samples is comparable. Therefore, we tend to interpret the slightly renormalized spin excitation energy and weakening of the Nd - Ni hybridization upon capping as a combination of effects such as interfacial reconstruction, crystalline defects, or lattice disorder \cite{Goodge2023,Yang2023,Zeng2021,Dantz2016,Hu2024}.

One of the important aspects is the atomic and electronic reconstruction at the SrTiO$_3$ - NdNiO$_2$ interface due to polar discontinuity \cite{Goodge2023,Yang2023,Zeng2021,Mundy2014,Geisler2020,He2020,Zhang2020,Bernardini2020,Ohtomo2004}. However, how the SrTiO$_3$ - NdNiO$_2$ interface affects the electronic properties of the inner bulk-like NdNiO$_2$ layers upon capping is still underexplored. Previous work reported that an intermediate layer such as Nd(Ti,Ni)O$_3$ forms, which increases the topotactic reduction energy \cite{Goodge2023}. Therefore, apical oxygen can be present at the interface shifting the Ni atoms renormalizing the Ni-Nd and Ni-O orbital hybridization \cite{Goodge2023}.~The interfacial atomic reconstruction has been proved to affect the electronic structure of nickelates for about 5 - 6 layers ($\sim$ 2 nm) \cite{Goodge2023,Yang2023}. Since the x-ray penetration depth (hundreds of nanometers) is much larger than the film thickness, the RIXS signal related to atomic and electronic reconstruction is about 20\% volume of our measured films ($c$ = 3.31 $\AA$).

To better understand how the interface renormalization affects the electronic properties of the bulk-like NiO$_2$ layers, we compute the atom- and layer-resolved density of states (DOS) of NdNiO$_2$ - cap and NdNiO$_2$ - uncap.~We calculate the most thermodynamically stable SrO-type of the interface (Supplementary Fig.~9) \cite{Bernardini2020} and the calculation considers five NiO$_2$ layers. Figure \ref{DFT calculation} displays the DOS of the Ni 3$d$, Nd 5$d$, and O 2$p$ orbitals of the NdNiO$_2$ - cap and NdNiO$_2$ - uncap. As expected, the most significant effects of ``capping'' manifest in the interfacial layer (layer 1), where the charge-transfer energy between Ni 3$d$ and O 2$p$ states is substantially enhanced compared to the ``uncapped'' scenario. Remarkably, the electronic renormalization from the interface also influences the inner NiO$_2$ layers. %While the bulk-like layers (layer 3) exhibit qualitative similarities in both scenarios, subtle quantitative distinctions emerge in the charge transfers between Ni 3$d$ - O 2$p$ and Ni 3$d$ - Nd 5$d$ states. 
The calculations with ``capping'' capture the decreased hybridization between the Ni 3$d$ and Nd 5$d$ states, evident in a larger Ni - Nd charge transfer energy ($\Delta_{\mathrm{Nd}}$) consistent with experimental results [Fig.~\ref{RIXS geometry}(c)]. This reveals that ``capping" can indeed affect the electronic properties of NdNiO$_2$ and the effects are not limited to the SrTiO$_3$-NdNiO$_2$ interface.

The magnetic behavior could also be rationalized by the interface effects.~As the Ni-Nd and Ni-O orbital hybridization decreases upon ``capping", less Ni$^{2+}$ states are present. This reduces the effective self-hole doping in the nickel site which hardens the spin excitations for the NdNiO$_2$ - cap. However, although the electronic renormalization can propagate to 5-6 NiO$_2$ layers into the film, the difference of the Ni-Nd hybridization between the NdNiO$_2$ - cap and NdNiO$_2$ - uncap is overall small since our film thickness is 10 nm which is larger than the length scale of the interface propagation effects. This is consistent with a minor renormalization of the spin excitation energy due to capping [Fig. \ref{spin dispersions}(a)].

%the relative difference of the Ni$^{2+}$ concentration between the capped and uncapped NdNiO$_2$ is small. The calculated $\Delta_{CT}$ and $\Delta_{Nd}$ overall increase about 4 $\sim$ 7\% and 6 $\sim$ 16\% in the bulk-like layers after capping, respectively, leading to a total 10 $\sim$ 23\% reduction of the effective hole-doping in the capped case. The $\Delta_{CT}$ of the interfacial layer increases about 40\% in the capped NdNiO$_2$. For the five-layer structure considered in Fig. \ref{DFT calculation} (two interfacial layers and three bulk-like layers), the expectation value of the effective hole-doping for capped NdNiO$_2$ is about 20 $\sim$ 25\% less than the uncapped case. This will contribute to $\approx$ 5\% enhancement in the magnon bandwidth considering the propagation of interface effects over the entire sample thickness (10 nm). The calculated volume fraction being affected by the interface reconstruction is qualitatively consistent with our extracted exchange constants ($\approx$ 7\%) from the spin excitation dispersion (Fig. 3).

Although the physical picture of the electronic renormalization induced by the interface is phenomenologically consistent with our data, we cannot fully disregard the effects from lattice or crystalline disorder (not the impurity phases like Nd$_3$Ni$_3$O$_7$ or Nd$_3$Ni$_3$O$_8$) due to the common challenges of synthesizing infinite-layer nickelate films \cite{Goodge2023,Raji2023}. %\textcolor{blue}{Indeed, the spectral weight of the spin excitations in the uncapped NdNiO$_2$ reduces around 20\% compared to the capped case [Fig. \ref{spin dispersions}(d)], which is a relatively large change compared to the renormalization of the magnon bandwidth. Furthermore, in 9 nm NdNiO$_2$, the magnon spectral weight increases about 15\% compared to the 10 nm film, comparable to the difference of magnon spectral weight between the capped and uncapped NdNiO$_2$ [Fig. \ref{spin dispersions}(d)]. This implies that the variation of the magnon spectral weight is within the sample-to-sample variations and potentially affected by other effects such as point defects and crystalline disorder.} 
Previous RIXS studies on infinite-layer cuprates (SrCuO$_2$)$_n$/(SrTiO$_3$)$_2$ suggested that the spin excitations can be renormalized due to the reorientation of the copper-oxygen plaquettes below the critical thickness caused by a relaxation of the polar electrostatic energy \cite{Dantz2016}.~The disordered states form scattering points for spin excitations, which quench the spectral weight of the spin excitations but do not change their overall bandwidth \cite{Dantz2016}. Our data could also be captured within this picture. However, it remains difficult to disentangle the impact of crystalline defects, strain relaxation, oxygen vacancy, cation structure and lattice disorder as they can vary from sample to sample which deserves future dedicated investigation.

In summary, we investigated the capping effects on the electronic spectra of infinite-layer nickelates.~In Nd$_{1-x}$Sr$_x$NiO$_2$ ($x$ = 0, 0.2), strong orbital polarization and spin excitations are present regardless of capping. We specifically identified that capping slightly hardens the spin excitation and weakens the Ni - Nd orbital hybridization. While our results can be phenomenologically rationalized by a combination of effects such as interface renormalization, crystal defects and lattice disorder, the x-ray data proves the intrinsic robustness of spin excitations in the phase diagram of infinite layer nickelates similarly to cuprates.

%\textcolor{blue}{While we cannot giv

%We specifically identified capping-induced effects on the spin and the Ni - Nd charge transfer excitations and rationalize the observed results by a combination of interfacial reconstruction as captured from our DFT calculations and lattice disorder and defects.} This evidence identifies the intrinsic physics of infinite layer nickelates and the role of interfaces.

\section{Acknowledgements} 

We thank Dr. Mark Dean and Dr. John Tranquada for fruitful discussions. Work at Brookhaven National Laboratory was supported by the DOE Office of Science under Contract No. DE-SC0012704. This work was supported by the Laboratory Directed Research and Development project of  Brookhaven National Laboratory No. 21-037. This work was supported by the U.S. Department of Energy (DOE) Office of Science, Early Career Research Program.  This research used beamline 2-ID SIX and 23-ID-1 CSX of the National Synchrotron Light Source II, a U.S. Department of Energy (DOE) Office of Science User Facility operated for the DOE Office of Science by Brookhaven National Laboratory under Contract No. DE-SC0012704. We also acknowledge resources made available through BNL/LDRD No. 19-013. The work at IOP is supported by the National Natural Science Foundation of China (Grant Nos. 12074411 and 11888101), Strategic Priority Research Program (B) of the Chinese Academy of Sciences (Grant No. XDB25000000). Y.N. would like to acknowledge the funding support from the National Natural Science Foundation of China (Grant Nos. 11861161004 and 11774153). ASB and AC acknowledge the LANEF chair of excellence. HL was supported by NSF Grant No. DMR 2045826.

\vspace{0.1in}

{\bf Competing interest:} The authors declare no competing interests.

\section{End Matter}

\subsection{High-quality of NdNiO$_2$ - uncap films}

%Add the thickness dependence, doping dependence, error bar analysis, and XAS at O-K and Ni L edges comparing with other samples, and the table shows different J values, thickness and lattice parameters.

\begin{figure}[t!]
\centering
\includegraphics[width=3.4in]{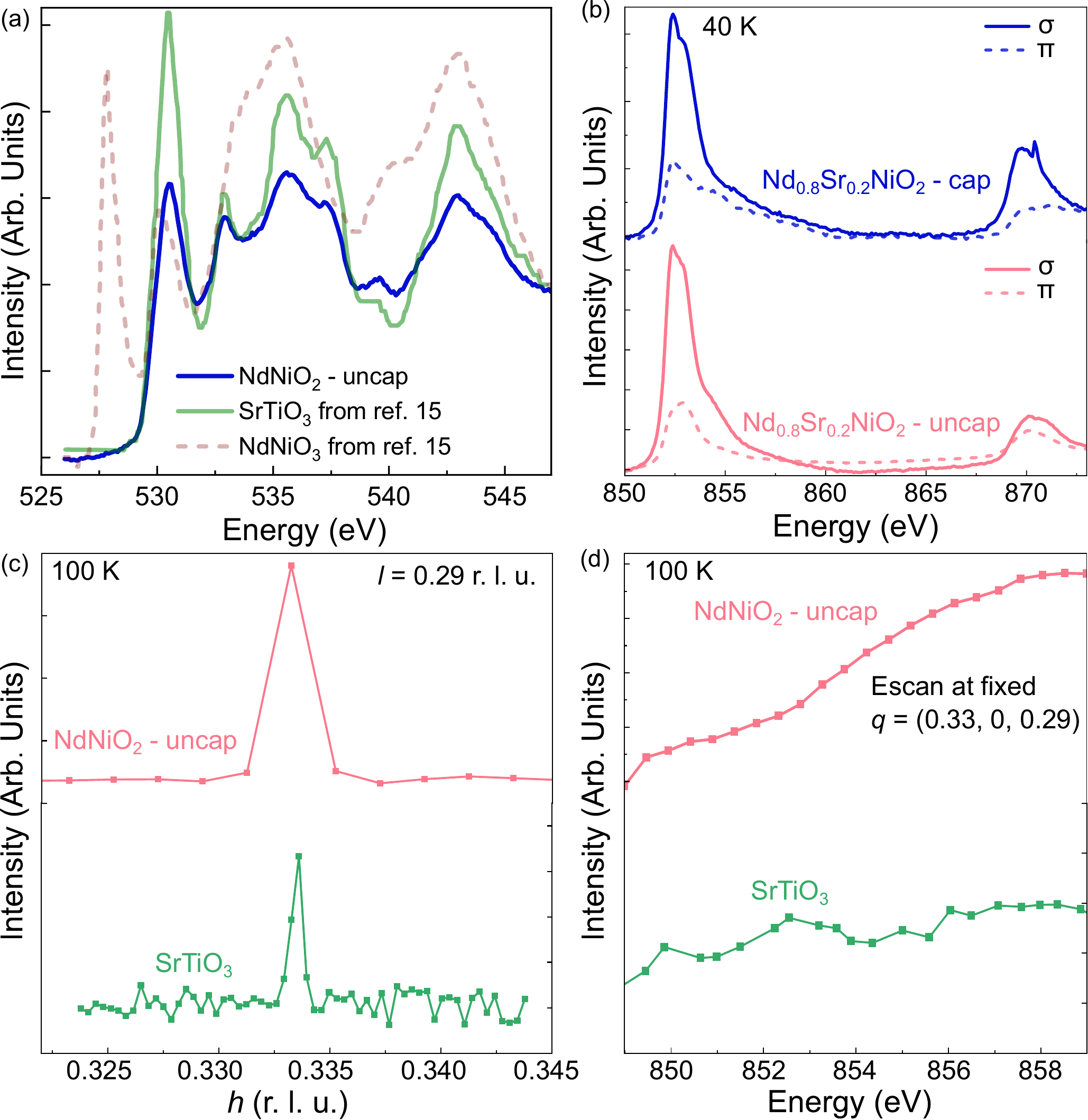}
\caption{(a) XAS comparison at the O-$K$ edge of the NdNiO$_2$ - uncap, the SrTiO$_3$, and perovskite NdNiO$_3$. The data of SrTiO$_3$ and NdNiO$_3$ grown on the SrTiO$_3$ substrate are reproduced from Ref. \citenum{Parzyck2023}. (b) XAS linear dichroism of the superconducting Nd$_{0.8}$Sr$_{0.2}$NiO$_2$ - cap/uncap at the Ni $L_3$ and $L_2$ edge. The sharp feature at about 870 eV in the $\sigma$ - polarized XAS of the Nd$_{0.8}$Sr$_{0.2}$NiO$_2$ - cap is a spike of the detector. (c) The Resonant x-ray diffraction (RXRD) scan along the ($h$, 0, 0.29) direction of the NdNiO$_2$ - uncap and the bare SrTiO$_3$ substrate at $E_i$ = 854 eV. (d) RXRD intensity as a function of incident x-ray energy of NdNiO$_2$ - uncap and SrTiO$_3$ at fixed $q$ = (0.33, 0, 0.29) r.~l.~u. across the Ni $L_3$ edge.~The background fluorescence was subtracted. The RXRD measurements were taken at 100 K.} 
\label{O-K edge}
\end{figure}

XAS at the O-$K$ edge is sensitive to the Nd$_3$Ni$_3$O$_7$ or Nd$_3$Ni$_3$O$_8$ impurity phases and we therefore use it to establish the quality of our samples \cite{Parzyck2023,Raji2023}. Figure \ref{O-K edge}(a) shows the XAS of the NdNiO$_2$ - uncap at the O-$K$ edge.  Despite small intensity variation, the spectrum displays nearly identical peaks to SrTiO$_3$ without any pre-edge peak due to the Ni-O hybridization as in NdNiO$_3$ \cite{Hepting2020,Parzyck2023}. This pre-edge peak is also observed in partially reduced films with reduced intensity compared to NdNiO$_3$ \cite{Raji2023}. This result proves that our film is fully or mostly reduced.

Figure \ref{O-K edge}(b) shows the XAS linear dichroism at the Ni $L_3$/$L_2$ edges of the superconducting Nd$_{0.8}$Sr$_{0.2}$NiO$_2$ - cap/uncap. Both capped and uncapped samples display strong dichroism at the Ni $L_3$ and $L_2$ edges, demonstrating a strong anisotropic orbital polarization in Sr-doped NdNiO$_2$. The strong XAS linear dichroism of parent and doped NdNiO$_2$ regardless of capping demonstrates the high quality of our entire sample sets and the minimal self-doping effect.

To further prove the quality of our NdNiO$_2$ - uncap, we perform resonant x-ray diffraction (RXRD) on NdNiO$_2$ - uncap and on bare SrTiO$_3$. Figure \ref{O-K edge}(c) displays the RXRD intensity along the ($h$, 0, 0.29) direction of NdNiO$_2$ - uncap and SrTiO$_3$. In both cases, we observe a peak at $h$ = 0.33 which can be indexed with the Bragg peak of SrTiO$_3$ (101) emerging from the third harmonic contamination of the beamline \cite{Pelliciari2023}. This is proved by the lack of any resonant enhancement across the Ni $L$ edge displayed in Fig. \ref{O-K edge}(d). The lack of resonance corroborates that this peak is structural and that the indexation shows that it originates from the substrate. Minor differences of the (0.33, 0, 0.29) peak between the NdNiO$_2$ - uncap and SrTiO$_3$ are due to the different reflectivity caused by the presence of interfaces between the substrate and the film.

The resonant $q$ = (1/3, 0, $L$) peak detected in other work \cite{Parzyck2023} is not detected in our uncapped NdNiO$_2$. Because of the extremely strong intensity of the intermediate Nd$_3$Ni$_3$O$_7$/Nd$_3$Ni$_3$O$_8$ phase when on-resonance, the $q$ = (1/3, 0, $L$) peak is detectable even if small amounts of the partially reduced phase are present \cite{Parzyck2023}. Therefore, the absence of such a peak in our NdNiO$_2$ - uncap demonstrates that our samples are free of the impurity phases \cite{Parzyck2023,Raji2023}.

\subsection{Error bar estimation and sample-to-sample variation}
The exchange constants in the different nickelates are determined by calculating the smallest $\chi^2$ from different combination of $J_1$ and $J_2$ values using linear spin-wave theory. The $\chi^2$ is directly related to the variance between the theory and the experimental spin excitation dispersion:

\begin{equation}
\chi^2 = \sum_{q_i}^{q_f}\frac{ 
 [\epsilon_{q_i}^{exp} - \epsilon_{q_i}^{cal}]^2}{\epsilon_{q_i}^{cal}}
\end{equation}
where $q_i$ and $q_f$ refer to the transferred momenta near the antiferromagnetic zone center and zone boundary, respectively. $\epsilon_{q_i}^{exp}$ and $\epsilon_{q_i}^{cal}$ are the experimental and calculated spin excitations energies at a specific momentum point. We keep the same ratio of $J_1$/$J_2$ for both NdNiO$_2$ - cap and NdNiO$_2$ - uncap to further constrain the fitting. Furthermore, due to the large uncertainties caused by the overlap between phonon and spin excitations at $q_{\parallel}$ $<$ 0.2 r. l. u., we remove the last two $q$ points. Using this method, we obtain $J_1$ = 72.8 $\pm$ 4.3 meV and $J_2$ = -5.8 $\pm$ 2.0 meV for the 10 nm NdNiO$_2$ - cap, and $J_1$ = 67.0 $\pm$ 4.0 meV and $J_2$ = -5.2 $\pm$ 2.0 meV for the 10 nm NdNiO$_2$ - uncap.

\begin{figure}[t!]
\centering
\includegraphics[width=3.4in]{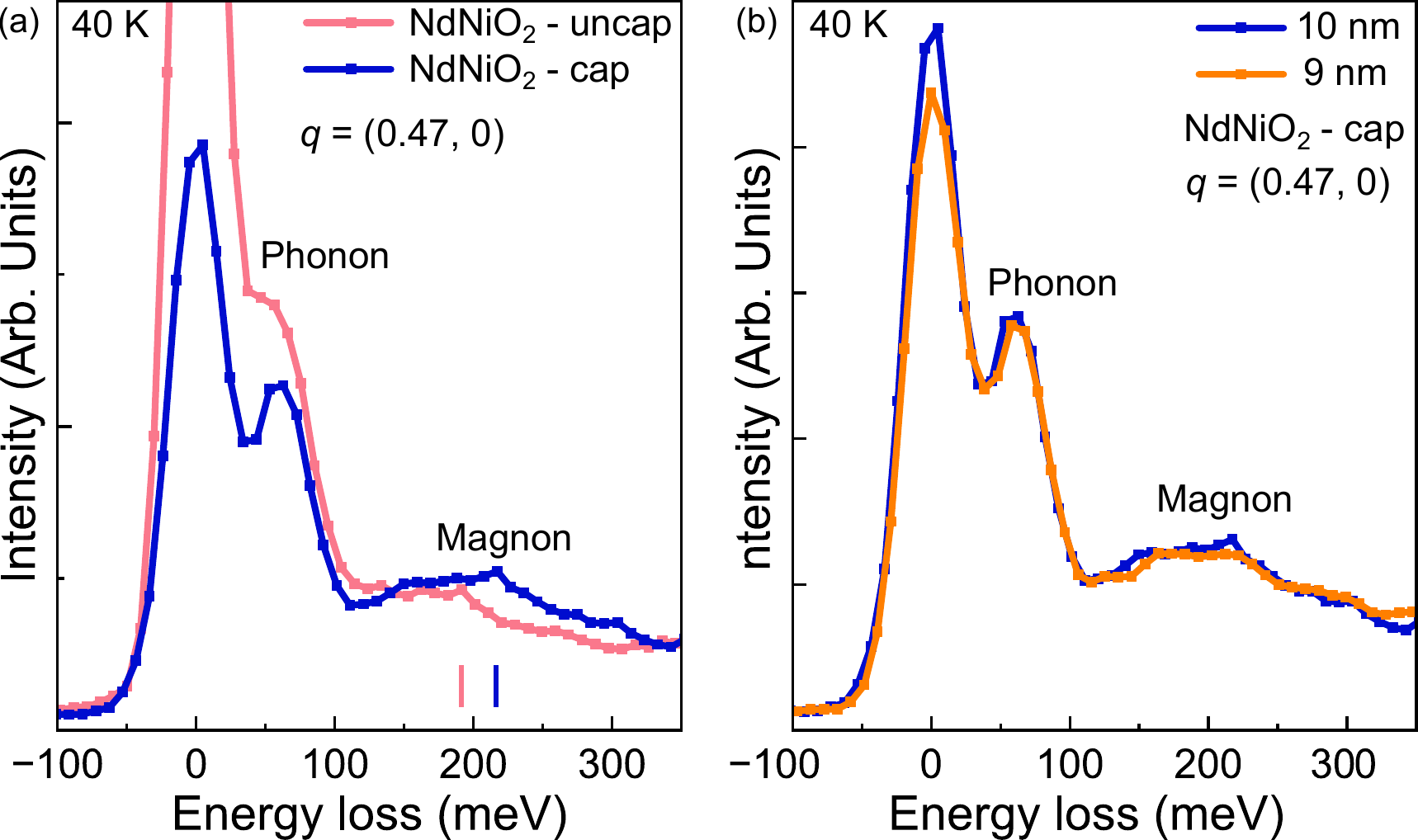}
\caption{(a) Comparison of the RIXS spectra of NdNiO$_2$ - cap and NdNiO$_2$ - uncap. The vertical lines indicate the spin excitation energies. (b) Comparison of the RIXS spectra of 9 and 10 nm NdNiO$_2$ - cap.} 
\label{9vs10nm}
\end{figure}

Considering the overall error bars, it is difficult to visualize the precise change of $J_1$ and $J_2$ due to capping. However, in Fig. \ref{spin dispersions} the bandwidth of spin excitation is systematically increased in capped samples for all momentum points. On the contrary, the spin excitation energies are nearly identical for 9 and 10 nm thick NdNiO$_2$ - cap. This proves that the capping induced spin excitation hardening is measurable, even if hardly quantifiable, and not due to sample-to-sample variations.

Figure \ref{9vs10nm}(a) shows the comparison of the raw RIXS spectra of NdNiO$_2$ - cap and NdNiO$_2$ - uncap at $q$ = (0.47,0) r.~l.~u., where spin excitations and phonon are well-separated. While the phonon remains at the same energy, capping hardens the spin excitation by about 20 meV. Figure \ref{9vs10nm}(b) displays the comparison of the raw RIXS spectra of 9 and 10 nm NdNiO$_2$ - cap. The spin excitations are nearly unchanged. This evidence shows that the capping-induced spin excitation hardening is significant compared to the level of sample-to-sample variations.

%The total error bars of $J_1$ and $J_2$ are estimated by comparing the difference between the two separate sets of exchange constants and also considering the uncertainties of the magnon bandwidth from the peak fitting at each momenta.

\bibliography{Nickelates_ref.bib,references_fromJP}

\end{document}